\documentclass[11pt]{article}
\pdfoutput=1

\usepackage{graphicx,epsf, epsfig, amssymb}
\usepackage{bm}
\usepackage{longtable}
\usepackage{color}
\usepackage{amsfonts,amsmath,amssymb,mathrsfs}

\usepackage{graphicx}
\usepackage{amssymb}
\usepackage{amsmath}
\usepackage{enumerate}
\usepackage{amsfonts}
\usepackage[section]{placeins}
\usepackage[colorlinks=true,linktocpage=true,linkcolor=blue,citecolor=blue]{hyperref}

\newcommand{\be}{\begin{equation}}
\newcommand{\ee}{\end{equation}}
\newcommand{\bea}{\begin{eqnarray}}
\newcommand{\eea}{\end{eqnarray}}

\newcommand{\dmoff}[1]{}

\def\beq{\begin{eqnarray}}
\def\eeq{\end{eqnarray}}

\def\ii{{{\rm i}\,}}
\def\mS{{\mathbb S}}
\def\mV{{\mathbb V}}
\def\mT{{\mathbb T}}
\def\mZ{{\mathbb Z}}

\newcommand{\beqa}{\begin{eqnarray}}
\newcommand{\eeqa}{\end{eqnarray}}

\newcommand{\Rmnum}[1]{\expandafter\@slowromancap\romannumeral #1@}

\numberwithin{equation}{section} 

\begin{document}

\pagestyle{plain}
\setcounter{page}{1}

\begin{titlepage}

\begin{center}
\vspace*{-1cm} \today 

\vskip 2.0cm

{\huge {\bf Superradiant instability of the Kerr brane}}

\vskip 14mm

{\large  {\bf Akihiro Ishibashi,$^{1}$ Paolo Pani,$^{2,3}$ Leonardo Gualtieri,$^{2}$ Vitor Cardoso,$^{3,4}$}}

\vspace{0.5 cm}


${}^1$ {\it Department of Physics, Kinki University, Higashi-Osaka 577-8502, Japan}

\medskip
  
${}^2$ {\it Dipartimento di Fisica, Universit\`a di Roma ``La Sapienza'' \& Sezione INFN Roma1, P.A. Moro 5, 00185, Roma, Italy.}

\medskip

${}^3$ {\it CENTRA, Departamento de F\'{\i}sica, Instituto Superior T\'ecnico, Universidade de Lisboa - UL,
Av.~Rovisco Pais 1, 1049 Lisboa, Portugal.}

\medskip

${}^4$ {\it Perimeter Institute for Theoretical Physics,
 Waterloo, Ontario N2J 2W9, Canada.}

\vskip 0.5cm

{\tt  akihiro@phys.kindai.ac.jp, \, paolo.pani@roma1.infn.it, \, leonardo.gualtieri@roma1.infn.it, \,
  vitor.cardoso@tecnico.ulisboa.pt}

\vspace{5mm}

{\bf Abstract}
\end{center}
 \noindent
We consider linear gravitational perturbations of the Kerr brane, an exact solution of vacuum Einstein's equations in dimensions higher than four and a low-energy solution of string theory. Decomposing the perturbations in tensor harmonics of the transverse Ricci-flat space, we show that tensor- and vector-type  
metric perturbations of the Kerr brane satisfy respectively a massive Klein-Gordon equation and a Proca equation on the four-dimensional Kerr space, where the mass term is proportional to the eigenvalue of the harmonics. Massive bosonic fields trigger a well-known superradiant instability on a Kerr black hole. We thus establish that Kerr branes in dimensions $D\geq6$ are gravitationally unstable due to superradiance. These solutions are also unstable against the Gregory-Laflamme instability and we discuss the conditions for either instability to occur and their rather different nature. When the transverse dimensions are compactified and much smaller than the Kerr horizon, only the superradiant instability is present, with a time scale much longer than the dynamical time scale. Our formalism can be also used to discuss other types of higher-dimensional black objects, taking advantage of recent progress in studying linear perturbations of four-dimensional black holes.

\noindent


\end{titlepage}

\tableofcontents

\section{Introduction}

In recent years, higher-dimensional black holes have come to play an essential 
role in understanding fundamental theories --~not only of gravitation 
but also of various different areas in physics~-- 
thanks to the celebrated gauge-gravity duality. 
It is by now well known that --~even in purely vacuum Einstein gravity~-- 
there exists a large variety of higher dimensional black objects, 
whose classification is still underway. 
Among these solutions, an interesting class is that of black branes, i.e., extended black 
objects taking the form of a direct product of a ($4$ or higher 
dimensional) black hole and some transverse, homogeneous extra-dimensions, 
which naturally arise in string theory.

It is clearly important to study the stability 
of such higher dimensional black objects. 
A peculiar feature of the stability problem of black branes is 
the Gregory-Laflamme (GL) instability \cite{Gregory:1994bj,Gregory:1993vy}, which is attributed in an essential way 
to the existence of uniform transverse-dimensions. Roughly speaking, this instability occurs when the length scale of compactification along the transverse 
dimensions is larger than the Schwarzschild radius.
If the black hole is rotating, one can also expect another type of 
instability --~i.e., superradiant (SR) instability~-- to occur. 
The latter instability is attributed to the combination of superradiant amplification~\cite{Teukolsky:1974yv} of 
incident waves by the black-hole rotation and a certain reflection mechanism 
that makes the scattered waves go back toward 
the hole repeatedly~[see e.g., Ref.~\cite{Brito:2015oca} and references 
therein]. Several different reflection mechanisms have been considered: for example, 
a ``mirror'' surrounding the hole imposed as a boundary condition~\cite{Press:1972zz,Cardoso:2004nk,Herdeiro:2013pia,Hod:2013fvl,Degollado:2013bha,Li:2014fna},  
the mass of perturbation fields~\cite{Damour:1976kh,Detweiler:1980uk,Dolan:2007mj,Pani:2012vp,Pani:2012bp,Witek:2012tr,Brito:2013wya,Yoshino:2015nsa}, nonminimal couplings~\cite{Pani:2013hpa,Cardoso:2013fwa,Cardoso:2013opa}, and the spacetime curvature itself 
for asymptotically AdS black holes~\cite{Hawking:1999dp,Cardoso:2004hs,Cardoso:2006wa,Uchikata:2009zz,Cardoso:2013pza}. 

For black branes with rotation, a possible reflection mechanism is played 
by the effective Kaluza-Klein mass due to excitations along 
the transverse dimensions. 
As such, in contrast to the GL instability, the superradiant instability is 
expected to occur even when the length scale of the transverse 
dimensions is smaller than the horizon radius of the rotating hole.  
There have already been a number of studies on the instabilities 
of black branes; most of which, however, have focused on the GL instability 
of nonrotating black branes. 
For rotating black branes, we are naturally led to ask which one --~either the GL or the SR instability~-- becomes dominant 
and/or under what circumstances such instabilities become relevant, and we therefore 
have to address both the GL and SR instabilities.

The purpose of this paper is to study the superradiant instability of 
rotating black branes against gravitational perturbations. 
For concreteness, we focus on the {\it Kerr branes}, which are the direct 
product of the $4$-dimensional Kerr metric and the transverse, $n$-dimensional, Ricci 
flat extra-dimensions. The complete solution satisfies the $(4+n)$-dimensional 
vacuum Einstein equations. 
Since the Kerr metric is believed to accurately describe black holes 
in astrophysical circumstances, analyzing perturbations of the Kerr brane 
would also be interesting in the context of exploring evidence for extra dimensions in our observable universe by exploiting recent developments 
of ``precision black-hole physics'' \cite{Arvanitaki:2009fg,Arvanitaki:2010sy,McClintock:2009as,Brito:2015oca}. 
The effects of extra dimensions may appear as massive bosonic fields and 
trigger a superradiant instability on the $4$-dimensional part of a 
Kerr brane, i.e., the Kerr metric. Then, comparisons with astrophysical 
observations, such as spin measurements of massive black holes, 
may provide some important pieces of information about compactified 
extra dimensions.

Our strategy is as follows. We first note that metric perturbations of 
the Kerr brane in $D = 4+n \geq 7$ dimensions can in general be classified 
into three types according to their tensorial behavior on the $n$-dimensional 
transverse Ricci flat dimensions: 
the tensor-, vector-, and scalar-type in terminology of 
Ref.~\cite{Kodama:2000fa}.  
(For $n = 2$, only two classes exist: the vector- and scalar-type, whereas 
for $n = 1$, i.e., Kerr-string, only scalar-type perturbations can be defined\footnote{
It is claimed in Ref.~\cite{Kudoh:2006bp} that all types --~tensor, vector, 
and scalar~-- of perturbations have been considered for the GL instability of 
a static black string, i.e., in the $n=1$ case. Note however that the way of 
decomposition adopted in \cite{Kudoh:2006bp} is based on the spherical 
section of the static black hole and is completely different from 
the method adopted in the present work. }.) 
In the present paper, we will focus on tensor- ($n\geq 3$) and 
vector- ($n \geq 2$) type perturbations.  
Next, we show that the linearized Einstein equations for the tensor- 
and vector-type metric perturbations of the Kerr brane reduce, respectively, 
to equations of motion for a massive scalar and massive vector field on 
the $4$-dimensional Kerr spacetime, where the effective mass term is associated with the eigenvalue of harmonics along the $n$-dimensional space. 
Then, by exploiting already available results for the superradiant instability  in $4$-dimensions, we establish the gravitational superradiant instability of the $D\geq6$ Kerr branes. 
In $4$-dimensional Kerr spacetime, superradiant instabilities 
triggered by ultralight massive scalar and vector fields have recently 
been studied in Refs.~\cite{Dolan:2007mj,Pani:2012vp,Pani:2012bp,Witek:2012tr}, 
and has been used to impose strong constraints on ultralight bosonic fields 
predicted in, e.g., the string axiverse scenario~\cite{Arvanitaki:2009fg,Arvanitaki:2010sy,Brito:2015oca}. 
As the Kerr branes can also be unstable against the GL instability, 
we discuss the conditions for either --~GL or SR~-- instability 
to occur and their rather different nature.

It is worth mentioning that our method for dealing with the perturbations 
described above can be applied not only to the Kerr branes but also to a 
much wider class of black objects, including Myers-Perry black branes and 
Myers-Perry black holes. 
Indeed, by employing a similar decomposition of the metric perturbations, 
the tensor-type perturbations with respect to the $n=D-4$ section of 
the $D \geq 7$ singly rotating Myers-Perry metric and the associated 
superradiant instability have been previously studied in 
Refs.~\cite{Kodama:2009bf,Kanti:2009sn,Doukas:2009cx}. However, 
the {\em vector-type} metric perturbations for rotating black objects 
are analyzed for the first time in the present paper and are shown to cause a
stronger instability than tensor-type perturbations.    

In the next section we establish the notations by describing 
our methods and formulas for perturbing black branes. 
In section \ref{sec:instability}, we first review the GL instability of 
black branes, and then discuss the superradiant instability of the Kerr brane 
by using the results of Refs.~\cite{Dolan:2007mj,Pani:2012vp,Pani:2012bp,Witek:2012tr}. 
We then compare the two instabilities and show that, when the transverse 
dimensions are compactified and much smaller than the Kerr horizon, 
only the superradiant instability is present, with an associated time scale much longer than the dynamical time scale. 
Section \ref{sec:conclusions} is devoted to summary and discussions, 
where we also discuss how our present analysis can be extended to 
more general black-brane cases. 

\section{Linear perturbations of the Kerr brane}
\label{sec:formalism}

The Kerr brane is the direct product of the four-dimensional spacetime 
described by the Kerr metric and an $n$-dimensional Ricci-flat space ${\cal K}$,
\begin{equation}
  ds^2=g_{MN}dx^M dx^N=g^{\rm Kerr}_{ab}dy^ady^b+R^2\gamma_{ij}dz^idz^j \,,
\label{kerrbrane}
\end{equation}
where (for instance, in Boyer-Lindquist coordinates) $y^a=(t,r,\theta,\phi)$, $\gamma_{ij}$ is the metric of ${\cal K}$,
and $R$ is a constant with dimensions of a length. The Ricci-flatness of $\gamma_{ij}$ and the fact that $g^{\rm  Kerr}_{ab}$ is a solution of 
the four-dimensional vacuum Einstein's equations guarantee 
that the full metric $g_{MN}$ is a solution of the $D=(4+n)$-dimensional 
Einstein's equations in vacuum. 

\subsection{The Kodama-Ishibashi formalism}
Before discussing the gravitational perturbations of a Kerr brane, we will 
present the relevant formulas in a rather generic setup. 
We follow the formalism of~\cite{Kodama:2000fa,Kodama:2003jz} (see also the review~\cite{Ishibashi:2011ws}), hereafter
KI, which allows to study linear perturbations of a $(d+n)$-dimensional spacetime ${\cal M}={\cal N}\times{\cal K}$,
where the $d$-dimensional space ${\cal N}$ is endowed with a Lorentzian metric $g_{ab}(y)$, and ${\cal K}$ is an
$n$-dimensional Einstein space with metric $\gamma_{ij}(z)$, i.e., $\hat R_{ij}[\gamma]=(n-1)K\gamma_{ij}$; $\hat
R_{ij}$ is the Ricci tensor on ${\cal K}$ and $K=0,\pm1$. The metric of the full space, in the coordinates
$x^M=(y^a,z^i$), has the form
\begin{equation}
 g_{MN}dx^Mdx^N = g_{ab}(y)dy^ady^b + R^2(y)\gamma_{ij}(z)dz^idz^j  \,, 
\label{KImetric}
\end{equation}
where now $R(y)$ can be a generic function of the coordinates $y^a$.
We denote by $\nabla_M$, $\nabla_a$, $\hat D_i$ the covariant derivatives on ${\cal M}$, ${\cal N}$, ${\cal K}$,
respectively. We also define $\hat\Delta={\hat D}^i\hat D_i$. 
It is clear that the Kerr-brane metric (\ref{kerrbrane}) is included 
as a special case of the metric ansatz (\ref{KImetric}).

Metric perturbations $h_{NM}=\delta g_{MN}$ of $({\cal M},g)$ are decomposed in tensor harmonics on ${\cal K}$, i.e., in
scalar harmonics ($\mS$), transverse vector harmonics ($\mV_i$) and transverse-traceless tensor harmonics ($\mT_{ij}$),
that satisfy the following relations: 
\begin{eqnarray}
\hat\Delta\mS&=&-k_S^2\mS \label{defS}\\
\hat\Delta\mV_i&=&-k_V^2\mV_i \label{defV}\\
\hat\Delta\mT_{ij}&=&-k_T^2\mT_{ij} \label{defT}
\end{eqnarray}
and
\begin{equation}
\hat D_i\mV^i=0~~;~~\hat D_i\mT^i_{~j}=0~~;~~\mT^i_{~i}=0\,.
\end{equation}
Since we are interested in vector and tensor perturbations, which exist for $n\geq2$ and $n\geq3$, respectively,
hereafter we assume $n\ge3$ (although our results are qualitatively valid also for $n=2$, see conclusions).

Perturbations are also expanded in ``longitudinal'' harmonic functions:
\begin{eqnarray}
\mS_i&=&-\frac{1}{k_S}\hat D_i\mS\\
\mS_{ij}&=&\frac{1}{k_S^2}\hat D_i\hat D_j\mS+\frac{1}{n}\gamma_{ij}\mS\\
\mV_{ij}&=&-\frac{1}{2k_V}(\hat D_j\mV_{i}+\hat D_i\mV_{j})\,.
\end{eqnarray}
Eqns.~(\ref{defS})-(\ref{defT}) can also be expressed in terms of the Lichnerowicz operator which, in an Einstein
space, can be written as $\hat\Delta_L=-\hat\Delta+2nK$; if $-k^2$ is the eigenvalue of the Laplace operator
$\hat\Delta$, the eigenvalue of the Lichnerowicz operator is $k^2+2nK$. 

Metric perturbations are decomposed in scalar-type, vector-type and tensor-type perturbations, i.e.,
$h_{\mu\nu}=h^S_{\mu\nu}+h^V_{\mu\nu}+h^T_{\mu\nu}$, which are expanded in the sets of harmonics $(\mS,\mS_i,\mS_{ij})$,
$(\mV_i,\mV_{ij})$, and $(\mT_{ij})$, respectively; it turns out that these three groups of perturbations are decoupled
in Einstein's equations. In the following we shall focus on tensor-type and vector-type perturbations only.

Tensor-type perturbations are expanded in terms of tensor harmonics as follows:
\begin{equation}
 h^T_{ab}=0, ~~~h^T_{ai}=0, ~~~h^T_{ij}= 2R^2 \Phi_T \mT_{ij}\,.
\end{equation}
The scalar function $\Phi_T(y)$ on the $d$-dimensional spacetime ${\cal N}$ is 
gauge invariant with respect to an infinitesimal coordinate transformation. 
Linearized Einstein's equations in vacuum yield~\cite{Kodama:2000fa}: 
\begin{equation}
  \nabla^a\nabla_a \Phi_T + n \frac{\nabla^aR }{R}\nabla_a\Phi_T 
  - \frac{k_T^2+2K}{R^2} \Phi_T =0\,. \label{eqtensor}
\end{equation}

Vector-type perturbations are expanded in terms of vector harmonics as follows:
\begin{equation}
  h^V_{ab}=0, ~~~h^V_{ai}=Rf_a\mV_i, ~~~h^V_{ij}= 2R^2 H_T \mV_{ij}
\end{equation}
where $f_a$ and $H_T$ are functions of $y^a$. The quantity
\begin{equation}
A_a = \frac{f_a}{R}+ \frac{1}{k_V}\nabla_a H_T 
\end{equation}
is gauge invariant with respect to an infinitesimal coordinate transformation. 
Linearized Einstein's equations in vacuum
yield~\cite{Kodama:2000fa}:
\begin{equation}
 \frac{1}{R^{n+2}} \nabla^{a}\left(R^{n+2} F_{ab}\right)  - \frac{k_V^2-(n-1)K}{R^2}A_b =0 \,,
\label{eqvector}
\end{equation}
where $F_{ab} = \nabla_a A_b - \nabla_b A_a$, and also  
\begin{equation}
k_V\nabla_a(R^n A^a) = 0  \,. 
\label{condi:vec}
\end{equation}
Note that $A_a$ and $F_{ab}$ above are different from those defined 
in \cite{Kodama:2000fa}. Here we use $A_a$ to be reminiscent of 
a Proca field. One can find the corresponding formulas, Eq.~(33), (35) 
and (36) in \cite{Kodama:2000fa} by replacing $A_a \rightarrow F_a/R$.

\subsection{Vector and tensor gravitational perturbations of the Kerr brane}
The Kerr brane belongs to the set of spacetimes considered in the KI formalism. When $d=4$, ${\cal N}$ is the Kerr spacetime, $K=0$ (i.e., ${\cal K}$ is Ricci-flat) and $R(y)=R$ constant (i.e., the product ${\cal N}\times{\cal K}$ is
not warped), the metric (\ref{KImetric}) reduces to the $D=(4+n)$-dimensional Kerr brane metric (\ref{kerrbrane}).

In this case, since $R$ is a constant, the perturbation equations for tensor-type and vector-type perturbations,
Eqns.~(\ref{eqtensor}), (\ref{eqvector}), (\ref{condi:vec}) reduce to 
\begin{align}
 \nabla^a\nabla_a \Phi_T  - \frac{k_T^2}{R^2} \Phi_T &=0\,, \label{eqtensorkerr}\\
\nabla^{b} F_{ab}  - \frac{k_V^2}{R^2}A_a &=0 \,, 
\\
k_V \nabla_aA^a &= 0 \,. 
\end{align}
For $k_V^2,k_T^2>0$, these are the equations of a massive scalar field and of a massive vector field, i.e., the massive
Klein-Gordon and the Proca equations with the Lorenz condition, respectively, 
propagating on 
a four-dimensional Kerr background. The effective masses are proportional 
to the eigenvalues of the harmonics, namely 
\begin{equation}
\mu_T=\frac{k_T}{R}\,,\qquad \mu_V=\frac{k_V}{R}\,.\label{massestv}
\end{equation}
It is well known that these perturbations trigger a {\it superradiant instability} of the Kerr metric and therefore
admit unstable modes, as discussed in Sec.~\ref{subsec:SR} below (see Refs.~\cite{Detweiler:1980uk,Cardoso:2005vk,Dolan:2007mj,Pani:2012vp,Brito:2013wya} and also the recent review~\cite{Brito:2015oca}). 
Therefore, this establishes that Kerr brane is subject to the superradiant instability\footnote{A similar instability has been
  discussed in Refs.~\cite{Cardoso:2004zz,Cardoso:2005vk} for the case of probe massless scalar perturbations of the
  Kerr brane. The novelty of our approach consists in the fact that the instability directly arises from the
  gravitational sector in the form of effective massive Klein-Gordon and Proca equations. As we discuss below, in this
  case the instability time scale is also shorter than for a probe scalar field.}.

\section{Superradiant and Gregory-Laflamme instabilities of the Kerr brane}
\label{sec:instability}
It is well known that the Kerr brane, like all black strings and black branes, is also subject to the Gregory-Laflamme
(GL) instability~\cite{Gregory:1993vy,Gregory:1994bj} (see also Ref.~\cite{Kudoh:2006bp}, the
review~\cite{Harmark:2007md} and references therein).  In this section, we summarize and compare the main features of
the superradiant instability and of the GL instability of the Kerr brane.
\subsection{The Gregory-Laflamme instability of black strings and branes}\label{subsec:GL}

The GL instability is a linear instability in the directions transverse to the brane (those described by the Ricci-flat
space ${\cal K}$).
It has first been studied for nonrotating, neutral black strings and branes~\cite{Gregory:1993vy}, for which the space
${\cal N}$ is the $d$-dimensional Tangherlini~\cite{Tangherlini:1963bw} spacetime (which reduces, for $d=4$, to the
Schwarzschild spacetime); subsequently, it has been generalized to charged~\cite{Gregory:1994bj} and to
slowly-rotating~\cite{Yoo:2011vu} strings and branes.

The black string/brane manifold is given by the metric (\ref{KImetric}) with $R(y)=R$ constant, 
 \begin{equation}
ds^2  =g_{ab}(y)dy^ady^b+R^2\gamma_{ij}(z)dz^idz^j\,,
 \label{brane} 
 \end{equation}
on ${\cal M}={\cal N}\times{\cal K}$ with $\cal K$ taken to be 
a Ricci flat space, and $\cal N$ a $d$-dimensional black hole manifold. If $\cal N$ describes a static, spherically symmetric
black hole, $g_{ab}$ is the Tangherlini metric, i.e., 
\begin{equation}
  g_{ab}dy^a dy^b=-fdt^2+f^{-1}dr^2+r^2d\Omega^2_{d-2}   
\end{equation}
where $d\Omega_{d-2}^2$ is the line element on $S^{d-2}$, $f=1-(r_0/r)^{d-3}$, and $r_0$ is the horizon radius; in
$d=4$, $r_0=2M$, where $M$ is the black-hole mass. More generally, $\cal N$ can be any black-hole spacetime; in particular, when
$d=4$ and $g_{ab}=g_{ab}^{\rm Kerr}$, we have the Kerr brane~\eqref{kerrbrane}.

The GL instability affects perturbations invariant under $S^{d-2}$ rotations, with the form $h_{ai}=h_{ij}=0$ and
\begin{equation}
h_{ab}(t,r,z)=\tilde h_{ab}(r) \mS(z) e^{\Omega t} \,, 
\label{exppert} 
\end{equation}
where $h_{ab}(r)$ is invariant under $S^{d-2}$-rotations, and $\mS(z)$ are 
eigenfunctions of the Laplace operator $\hat\Delta$ on ${\cal K}$ 
defined by (\ref{defS}). 
Let us focus on stationary perturbations, i.e. those with $\Omega=0$ which, 
as we discuss below, describe the onset of the instability. 
With an appropriate gauge choice, the perturbation equations have the form 
\begin{equation}
\Delta_Lh_{ab}=\mS(z)\Delta_L^d\tilde h_{ab}-\tilde h_{ab}\hat\Delta \mS(z)=0 \,, 
\label{GLperteq}
\end{equation}
where $\Delta_L$ is the Lichnerowicz operator on ${\cal M}$, and $\Delta_L^d$ is the Lichnerowicz operator on the
$d$-dimensional manifold ${\cal N}$. Using Eq.~(\ref{defS}), 
the perturbation equation on ${\cal N}$ reads
\begin{equation}
  \Delta_L^d\tilde h_{ab}(r)=-k_{\Omega=0}^2\tilde h_{ab}(r) \,.\label{deigeneq}
\end{equation}
When ${\cal N}$ is the Tangherlini spacetime, this equation admits the eigenvalue
\begin{equation}
k_{\Omega=0}^2=\frac{k_c^2}{r_0^2}\,, \label{GLperteq2}
\end{equation}
where $k_c \approx 0.88$ for $d=4$, $k_c\approx 1.24$ for $d=5$ and so on~\cite{Harmark:2007md}~\footnote{We remark that
  the linearized Einstein's equations on ${\cal N}$ can be written as $\Delta_L^d\tilde h_{ab}(r)=0$, while
  Eq~(\ref{deigeneq}) is equivalent to the massive graviton equation on ${\cal N}$~\cite{Babichev:2013una}. The stable
  and unstable modes of this equation on a Kerr background have been discussed in detail in Ref.~\cite{Brito:2013yxa}.}.
Note that $k_{\Omega=0}=k_c/r_0$ is the wavenumber of the marginally stable mode and has dimensions of inverse length, whereas
$k_c$ is a numerical coefficient.  The wavelength of the marginally stable mode is then
$\lambda=\lambda_{GL}\equiv 2\pi r_0/k_c$. Gregory and Laflamme have shown that if a neutral mode --~solution of
Eq.~(\ref{GLperteq2})~-- exists, then for any wavenumber in the range $0<k<k_c/r_0$, i.e., for any wavelength
$\lambda>\lambda_{GL}$, there exists an unstable mode, with $\Omega>0$. Generally, the instability time scale is $\tau_{\rm
  GL}=\Omega^{-1}\sim r_0$~\cite{Harmark:2007md}, therefore the instability grows on a dynamical time scale.

If ${\cal K}$ is non-compact, Eq.~(\ref{defS}) 
admits a continuum spectrum of eigenvalues $k$, Eq.~(\ref{GLperteq2})
always has a solution, and a neutral mode with wavenumber $k_c/r_0$ and wavelength $\lambda_{GL}$ always exists.  If,
instead, ${\cal K}$ is a compact space, Eq.~(\ref{defS}) 
has a discrete set of eigenvalues, bounded from below by the critical
wave number $k_{\rm min}$, corresponding to the critical wavelength $2\pi/k_{\rm min}$. The neutral mode exists if
Eq.~(\ref{GLperteq2}) can be satisfied, i.e., if
\begin{equation}
 k_{\rm min}\le \frac{k_c}{r_0}\,.\label{GLcond}
\end{equation}
In the case of a black string compactified on a circle with length $L$, the critical wavenumber is $k_{\rm min}=2\pi/L$,
thus - if the condition (\ref{GLcond}) is satisfied, i.e. if $2\pi r_0/L\ge k_c$ - the critical wavelength is $L$, and
the instability sets in for modes with $\lambda\ge L$.  In the case of a torus $\mT^n$ \cite{Kol:2004pn}, the critical wave
vector is $k_{\rm min}=|\vec k_{\rm min}|$, where $\vec k_{\rm min}$ is the shortest vector in the reciprocal lattice of
$\mT^n$. If, for instance 
the torus is a simple product of $n$ circles with lengths $L_1\ge L_2\ge\dots\ge L_n$, $\vec k=2\pi(m_1/L_1,m_2/L_2,\dots)$ with $m_j\in\mZ/\{0\}$, then $k_{\rm min}=2\pi/L_1$, and the critical wavelength is $2\pi/k_{\rm min}=L_1$: the instability sets in
(if the condition (\ref{GLcond}) is satisfied) for modes whose wavelength is larger than the {\it largest} of the
circles composing the torus.

The case of a slowly-rotating black brane, in $d=4$, has been studied 
in \cite{Yoo:2011vu}, where the spacetime ${\cal N}$ is described by 
the Kerr metric in Boyer-Lindquist coordinates, expanded in dimensionless 
angular momentum $\tilde a=J/M^2$ up to ${\cal O}({\tilde a}^2)$, where 
$M$ and $J$ denote the mass and angular momentum of the Kerr black hole. 
Note that since $d=4$, $r_0=r_+=2M(1-\tilde{a}^2/2)+{\cal O}(\tilde{a}^4)$ is 
the horizon radius of the slowly-rotating Kerr solution. In this perturbative scheme, rotation only modifies the eigenvalue of the four-dimensional equation (\ref{deigeneq}), 
whereas the harmonic expansion of the perturbations~(\ref{exppert}) is unaffected. The modified eigenvalue of the neutral mode reads
\begin{equation}
 k_{\Omega=0}=\frac{k^{\rm rot}_c}{r_+}\,, \label{muGLrot}
\end{equation}
where\footnote{In the notation of \cite{Yoo:2011vu}, $\tilde a\rightarrow\epsilon$, $k^{rot}_c\rightarrow \mu r_+$, and
  $\gamma\rightarrow\mu_2^2/(4\mu_0^2)-1/4$. Note that in Figure~1 of \cite{Yoo:2011vu} the wavenumber eigenvalue is
  rescaled with $M$, while in this paper, to facilitate the comparison with the superradiant instability, it is
  rescaled with $r_+$.} 
\begin{equation}
 k_c^{\rm rot}=k_c(1+\gamma {\tilde a}^2)+{\cal O}({\tilde a}^4)\,,
\end{equation}
From Figure~1 of~\cite{Yoo:2011vu}, it is possible to estimate the value of the parameter $\gamma\simeq 1.45$. 
The rest of the discussion on GL instability is not affected by the rotation of the black string/brane, which -
at least, for slow rotation - only increases the value of the critical wavenumber, making the instability stronger.

\subsection{The superradiant instability}\label{subsec:SR}
In the presence of a massive bosonic field (such as a massive scalar or vector field), a Kerr black hole is
unstable~\cite{Damour:1976kh,Detweiler:1980uk,Cardoso:2005vk,Dolan:2007mj,Pani:2012bp,Brito:2013yxa} due to
\emph{superradiance} (see Ref.~\cite{Brito:2015oca} for an overview). In brief, the process relies on the scattering of
a monochromatic wave off a spinning black hole; the energy of the wave is amplified when its frequency satisfies the
condition
\begin{equation}
 0<\omega/m<\Omega_{\rm H}\,, \label{SRcond}
\end{equation}
where $|m|=0,1,\dots,\ell$ is the azimuthal number along the axis of rotation, $\ell=0,1,\dots$ is the harmonic index,
and $\Omega_{\rm H}=\tilde{a}/(2r_+)$ is the angular velocity of the horizon. This superradiant amplification can
trigger an instability whenever the amplified radiation is confined~\cite{Brito:2015oca}.

Massive bosonic perturbations support long-lived modes with a hydrogenic spectrum in frequency,
$\omega\sim\mu$, where $\mu$ is the mass of the field.
These modes satisfy the superradiant condition when
\begin{equation}
 \mu\lesssim m\Omega_{\rm H} =\frac{\tilde{a}m }{2 r_+}\,. \label{boundmass}
\end{equation}
In addition, the mass of the perturbation provides a potential barrier that can trap the superradiant modes near the
black hole. When the condition~\eqref{boundmass} is satisfied, rotational
energy is extracted from the horizon through superradiance, 
while the trapping due to the mass is necessary to ``keep the
extraction going'', thereby triggering the instability. 

In the limit $r_+ \mu\ll1$, the instability time scale of 
the fundamental unstable mode 
is~\cite{Detweiler:1980uk,Pani:2012vp,Pani:2012bp,Brito:2013yxa} 
\begin{equation}
  \tau_{\rm SR}\sim\frac{M}{\gamma_{S\ell}}\frac{\left(M
    \mu\right)^{-(4\ell+5+2S)}}{\tilde{a}m-2 r_+ \mu} \,,\label{tauSR}
\end{equation}
where $S=-s,-s+1,\ldots,s-1,s$ is the spin projection, $s$ being the spin of the bosonic field ($s=0,1,2$ for scalar,
vector and tensor fields, respectively), and $\gamma_{S\ell}$ is a numerical coefficient depending on $S$, $\ell$ and $m$.
In the opposite regime, when
$r_+ \mu\gg1$, the instability time scale is~\cite{Zouros:1979iw}
\begin{equation}
 \tau_{\rm SR}^{\rm WKB}\sim 10^7 e^{3.7 r_+\mu} r_+ \,, \label{tauWKB}
\end{equation}
for any type of field. A numerical analysis shows that the shortest instability time scale occurs for highly-spinning
black holes, low values of $\ell=m$, and in the cross-over region $r_+ \mu\sim1$. For a given rotation rate
and fixed value of $m$, the maximum instability occurs slightly before the condition~\eqref{boundmass} is
saturated. However, even in the most favorable case for the instability, the time scale is always much longer than the
dynamical time scale, i.e. $\tau_{\rm SR}\gg M$~\cite{Dolan:2007mj,Brito:2015oca}.
In the scalar case, the minimum instability time scale $\tau_{\rm SR}\sim 10^6 M$ occurs for $\ell=m=1$, $\mu M\sim 0.42$,
and near-extremal black holes~\cite{Dolan:2007mj}. For $\ell=m=2$, the strongest instability in the
scalar case occurs when $\mu M\sim 0.9$ and corresponds to $\tau_{\rm SR}\sim 10^7 M$. In the vector case, most
results were obtained in the slow-rotation limit~\cite{Pani:2012vp,Pani:2012bp} (cf. Ref.~\cite{Pani:2013pma} for a
review) and are valid to ${\cal O}(\tilde{a}^2)$, although fully-numerical results~\cite{Witek:2012tr} are in good
quantitative agreement with an extrapolation of the slow-rotation approximation. In this case, the strongest instability
occurs for $\mu M\sim 0.39$ for $\ell=m=1$ and corresponds to $\tau_{\rm SR}\sim 10^3 M$~\cite{Witek:2012tr},
in good agreement with the extrapolation discussed in Ref.~\cite{Pani:2012bp}.  As expected from the fact that
electromagnetic waves are more strongly amplified by black-hole superradiance, the minimum instability time scale of
Proca fields is much shorter than that for massive scalar perturbations~\cite{Pani:2012vp,Pani:2012bp}. The instability
time scale for Proca field with higher $\ell$ has not been investigated in detail, but it should agree with the WKB
result~\eqref{tauWKB}, because the latter is insensitive to the spin of the perturbation when $\ell,m\gg1$.

\subsection{Superradiant instability versus Gregory-Laflamme instability}
When $d=4$ and ${\cal N}$ is the Kerr spacetime, i.e. $g_{ab}=g_{ab}^{\rm Kerr}$, 
and ${\cal K}$ is Ricci-flat and $R(y)=R$ constant, 
the metric (\ref{KImetric}) reduces to the Kerr brane (\ref{kerrbrane}) 
as explained before. In this case $r_0=r_+$ coincides with the horizon 
location of the Kerr metric. 

The Kerr brane can be subject both to the GL instability and to the superradiant instability. These instabilities are of
very different nature. On the one hand, the perturbations corresponding to the GL instability depend on the directions
\emph{transverse} to the brane, $z^i$, but do not depend on the angular directions $(\theta,\phi)$ of Kerr spacetime.

These perturbations tend to fragment the brane into lower-dimensional objects, such as black holes~\cite{Lehner:2010pn},
similarly to the Rayleigh-Plateau instability of fluids~\cite{Cardoso:2006ks,Camps:2010br}.
On the other hand, the perturbations corresponding to the superradiant instability depend also on the
angular directions of the Kerr black hole. 
Because of the azimuthal dependence, such instability extracts angular momentum from the horizon, thus slowing the black
hole down~\cite{Brito:2015oca}.

It is therefore natural to ask in which conditions one or the other instability is present, and which one prevails in
the cases both instabilities are at work. Since the GL instability grows on a dynamical time scale $\tau_{\rm GL}\sim
M$~\cite{Harmark:2007md}, while the superradiant instability grows on a much longer time scale as discussed above, the
former instability prevails when both are present. This happens when the space ${\cal K}$ is
non-compact: as discussed in Sec.~\ref{subsec:GL}, in this case the GL instability always occurs.

Nonetheless, there are configurations in which the GL instability is absent and the sole process governing the dynamics
of the solution is the superradiant instability. This happens, for instance, when the space ${\cal K}$
is a {\it compact} (Ricci-flat) manifold.
For simplicity, we shall assume that ${\cal K}$ is a torus $\mT^n$, product of $n$ identical circles, each of them with
length $L$ (as discussed above, when the circles are different, it is the largest one that determines the onset of the
instability). 
The solutions of the Laplace equations on the torus are simply 
$\mS(z)\sim e^{\ii k_iz^i}$ with $k_i=2\pi m_i/L$ and 
$m_i\in\mZ/\{0\}$ ($i=1,\dots,n$). Therefore, the minimum eigenvalue is $k_{\rm min}=2\pi/L$.  The GL instability is present when $k_{\rm min}\le k_c^{\rm rot}/r_+$, i.e., when
\begin{equation}
  \frac{2\pi}{L}\le \frac{k_c^{\rm rot}}{r_+}\,. \label{condGL}
\end{equation}
The KI formalism is also very simple when the internal space is the torus $\mT^n$ (for simplicity, we consider the case in which all circles have
equal length $L$). In this case we have $R=L/(2\pi)$, $\gamma_{ij}=\delta_{ij}$, and the equations for scalar, vector and tensor
harmonics reduce to
\begin{equation}
[\hat\Delta+k_S^2]\mS=[\hat\Delta+k_V^2]\mV_i=[\hat\Delta+k_T^2]\mT_{ij}=0
\end{equation}
where $\hat\Delta=\delta^{ij}\partial_i\partial_j$, with eigenvalues $k_S=0,1,\dots$ and $k_V,k_T=1,2,\dots$, respectively. As discussed in Sec.~\ref{sec:formalism},
vector and tensor perturbations satisfy the equations of massive vector fields and massive scalar fields, respectively,
with masses $\mu=\mu_{T,V}=k_{T,V}/R$ [cf. Eq.~\eqref{massestv}]. This yields a lower bound on the mass,
\begin{equation}
\mu\ge\frac{1}{R}=\frac{2\pi}{L}\,.
\end{equation}
Combining the bound above with the superradiance condition~\eqref{boundmass} yields a necessary condition for the superradiant instability:
\begin{equation}
  \frac{2\pi}{L}\lesssim \frac{\tilde am}{2r_+}\,. \label{condSR}
\end{equation}
By comparing Eq.~\eqref{condSR} with Eq.~\eqref{condGL}, we conclude that Kerr branes such that
\begin{equation}
 k_c^{\rm rot}\le \frac{2\pi r_+}{L} \lesssim  \frac{\tilde{a} m}{2}\,, 
\label{finalcond}
\end{equation}
are unstable only against the superradiant instability. Interestingly, the upper limit of this condition is arbitrarily large, as $m$ can be any integer. However, when $m\gg 1$ the instability time scale is strongly suppressed~\cite{Zouros:1979iw,Dolan:2007mj,Brito:2015oca}.


As an example, let us consider a slowly-rotating Kerr brane with $\tilde a=0.2$. In this case $k_c^{\rm rot}\approx 0.93$, therefore the GL instability is present when
\begin{equation}
\frac{2\pi r_+}{L}\lesssim 0.93 \,, \label{cond1}
\end{equation}
whereas the superradiant instability is present when
\begin{equation}
\frac{2\pi r_+}{L}\lesssim\frac{\tilde am}{2}=0.1m\,.\label{cond2}
\end{equation}
When condition (\ref{cond1}) is violated, condition (\ref{cond2}) can still be satisfied, as long as
$m\gtrsim10$. 

To summarize, while the superradiant instability is always present (there is always a value of $m$ large enough to satisfy
the upper bound in Eq.~\eqref{finalcond}), the GL instability is only present if the compactified dimensions are sufficiently large. For smaller circles, the superradiant instability is
the only one to occur. On the other hand, in this case the superradiant instability is associated with modes of large $m$, and thus it is strongly suppressed. These results have been derived for small values of the rotation rate, but it is reasonable to expect that the qualitative description holds also for rapidly-spinning black holes. 
By extrapolating Eq.~\eqref{finalcond} to the extremal limit, we find a lower bound on $m$ which is necessary to have the superradiant instability but not the GL instability,
\begin{equation}
 m\gtrsim \frac{2 k_c^{\rm rot}}{\tilde{a}}\sim 4.3\,.
\end{equation}
where in the last step we have extrapolated to $\tilde{a}\sim 1$.
From Eqs.~\eqref{tauWKB} and~\eqref{boundmass}, the minimum instability time scale in this case is approximately
\begin{equation}
 \tau_{\rm SR}^{\rm WKB}\sim 10^{11} M\,.
\end{equation}

\section{Concluding remarks}
\label{sec:conclusions}
We have shown that Kerr branes in $D \geq 6$ are gravitationally unstable. 
For this purpose, we have (i) considered the tensor- and vector-type metric 
perturbations with respect to the $n=D-4$ dimensional Ricci flat space, 
i.e., the brane direction, and 
(ii) reduced the linearized Einstein equations to the equations for 
a massive scalar and a Proca field, and then (iii) compared with already 
available results for the superradiant instability of 
$4$-dimensional Kerr black holes against massive scalar and Proca fields. 
Note that, in order to deal with both the tensor and vector perturbations 
simultaneously, we have proceeded our arguments mainly 
in the $D \geq 7$ ($n \geq 3$) case as the tensor-type perturbations 
exist only when $n \geq 3$. However, our results concerning the vector-type 
perturbations also hold for the $D=6$ ($n = 2$) case. 
We have found that the instability is stronger for the vector perturbations 
than the tensor perturbations, as shown in (\ref{tauSR}). 
However, this is only true for low $m$ and $\ell$, 
as in the eikonal limit, Eq.~\eqref{tauWKB} should be valid for 
all perturbations.
When the Ricci flat brane direction, $\cal K$, is non-compact, 
the GL instability always occurs and grows on a dynamical time scale, 
and eventually prevails over the superradiant instability, whose 
time scale is much longer than the dynamical time scale.  
However, when the brane direction $\cal K$ is compactified in 
sufficiently small length scale, the GL instability becomes absent. 
We have compared the two instabilities, and derived 
the condition (\ref{finalcond}) under which Kerr branes become unstable only 
due to superradiant instability.

In the present paper, we focused on the tensor- and vector-type perturbations 
since for these two cases we could take advantage of already available results \cite{Dolan:2007mj,Pani:2012vp,Pani:2012bp,Witek:2012tr} 
for the superradiant instability in $4$-dimensional Kerr spacetimes. 
We can also apply our formalism to the scalar-type gravitational 
perturbations for Kerr-strings/branes with $n\geq 1$, in which 
the linearized Einstein equations reduce to equations for 
a massive tensor field on $4$-dimensional Kerr spacetime. 
Although the equations of motion become much more complicated, 
by applying the slow-rotation approximation~\cite{Pani:2012vp,Pani:2012bp,Pani:2013pma}, 
one should be able to simplify the equations of motion for the massive tensor 
field on the Kerr spacetime (see Ref.~\cite{Brito:2013wya}) and study its superradiant instability. 
This analysis is left for future work.

Our formalism can also be used to study perturbations of 
other types of higher-dimensional black objects. 
For example, the $4$-dimensional Kerr subspace of our Kerr brane can be 
replaced with a $d$-dimensional Myers-Perry metric, thereby obtaining 
$D= d+n$ Myers-Perry branes (MP-branes). 
Then, as in the Kerr brane case examined here, tensor- and vector-type metric 
perturbations of the MP-branes reduce to a massive scalar and Proca fields 
on the $d$-dimensional MP black-hole metric. This type of reduction would 
help us to compute, e.g., quasi-normal modes of $(d+n)$-dimensional MP branes 
by exploiting quasi-normal mode analysis on $d$-dimensional MP black hole. 
We can also apply our method to analyze perturbations of the MP black hole 
itself. It is known that when $D$-dimensional MP black hole admits only 
a single angular momentum, its metric is written as the warped product form 
${\cal M} = {\cal N} \times {\cal K}$ of the metric ansatz (\ref{KImetric}). 
More concretely, let us take $y^a=(t,r,\theta, \phi)$ as the coordinates 
on $4$-dimensional spacetime $\cal N$, and set the $n= (D-4)$-dimensional 
metric on $\cal K$ as the spherical metric, 
$\gamma_{ij}dz^idz^j= d\Omega^2_{(n)}$ times the square of the warp factor 
$R(y)= r \cos \theta$. Then the $D=(4+n)$-dimensional MP black hole with a single 
rotation parameter is given by setting the $4$-dimensional metric $g_{ab}$ on $\cal N$ as 
the following Kerr-like metric: 
\begin{eqnarray}
 g_{ab}dy^ady^b &=& - dt^2 + \sin^2\theta (r^2+a^2)d\phi^2 
                  + \frac{\mu}{r^{n-1}\rho^2}(dt-a\sin^2  \theta d\phi)^2 
\nonumber \\
              &&{} + \frac{\rho^2}{r^2+a^2-\mu r^{1-n}}dr^2 
                   + \rho^2 d\theta^2 \,, 
\label{metric:Kerr:type}
\end{eqnarray}
where $\mu$ and $a$ denote the mass and spin parameters, respectively, 
and $\rho^2 := r^2 + a^2 \cos^2 \theta$. 
As mentioned before, when $D \geq 7$, one can consider 
tensor-type perturbations with respect to the $n (\geq 3)$-sphere, 
for which the linearized Einstein equations reduce to 
a massive Klein-Gordon equation on the $4$-dimensional 
Kerr-like spacetime (\ref{metric:Kerr:type}). This case has been studied 
previously in Refs.~\cite{Kanti:2009sn,Doukas:2009cx}. When $D = 6$, there are 
no tensor perturbations but instead, the vector and scalar-type perturbations 
become relevant. 
Then, problems of computing, e.g., quasi-normal modes of MP black holes (see e.g. Ref.~\cite{Dias:2014eua})
for these types of metric perturbations would reduce to the problem of 
analyzing massive vector and $2$nd-rank tensor fields on the $4$-dimensional 
Kerr-like spacetime $({\cal N}, g_{ab})$.

\section*{Acknowledgments}
We would like to thank the authors of reference~\cite{Yoo:2011vu}, 
especially Chul-Moon Yoo, for discussions and for sharing with us detailed data of their analysis. 
We thank the Yukawa Institute for Theoretical Physics at Kyoto University for hospitality during the YITP-T- 14-1
workshop on ``Holographic vistas on Gravity and Strings.''
P.P. was supported by the European Community through
the Intra-European Marie Curie contract AstroGRAphy-2013-623439 and by FCT-Portugal through the project IF/00293/2013.
V.C. acknowledges financial support provided under the European
Union's FP7 ERC Starting Grant ``The dynamics of black holes: testing
the limits of Einstein's theory'' grant agreement no. DyBHo--256667,
and H2020 ERC Consolidator Grant ``Matter and strong-field gravity: 
New frontiers in Einstein's theory'' grant agreement no. MaGRaTh--646597.
A.I. was supported in part by JSPS KAKENHI Grant Number 15K05092 
and 23740200. 
This research was supported in part by the Perimeter Institute for Theoretical Physics. 
Research at Perimeter Institute is supported by the Government of Canada through 
Industry Canada and by the Province of Ontario through the Ministry of Economic Development 
$\&$ Innovation.
This work was supported by the NRHEP 295189 FP7-PEOPLE-2011-IRSES Grant.

\bibliographystyle{myutphys}
\bibliography{biblio}
\end{document}